\documentclass[twocolumn,showpacs,preprintnumbers,amsmath,amssymb,prl]{revtex4}

\usepackage{epsfig}

\newcommand{\phiR}{\varphi_r}

\newcommand{\EFLR}{E_{F,L/R}}

\begin{document}

\title{Interaction-induced beats of Friedel oscillations in quantum wires}
\author{D.~F.~Urban${}^1$ and A.~Komnik${}^{1,2}$}
\affiliation{${}^1$Physikalisches Institut,
Albert-Ludwigs-Universit\"at, Hermann-Herder-Str. 3, D-79104
Freiburg, Germany \\
${}^2$Institut f\"ur Theoretische Physik, Universit\"at
Heidelberg, Philosophenweg 19, D-69120 Heidelberg, Germany}
\date{\today}

\begin{abstract}
 We analyze the spectrum of electron density oscillations in an
 interacting one-dimensional electron system with an impurity.
 The system's inhomogeneity is characterized by different values
 of Fermi wave vectors $k_F=k_{L/R}$ on left/right side of the
 scatterer, leading to a Landauer dipole formation.
 We demonstrate, that while in the noninteracting system
 the Friedel oscillations possess only one periodicity related to
 the local $k_F$, say $k_L$ on the left side, the interplay of the
 interactions and the Landauer dipole generates an additional
 peak in the spectrum of density oscillations at the counterpart $k_R$.
 Being only present in correlated systems, the position and shape
 of this spectral feature, which in coordinate space
 is observable as a beating pattern in the Friedel oscillations,
 reveals many important details about the
 nature of interactions. Thus it has a potential to become an
 investigation tool in condensed matter physics.
\end{abstract}


\maketitle


The role of electron-electron interactions is decisive in a large
number of condensed matter phenomena. The correlations are not
only responsible for such spectacular effects as the fractional
quantum Hall effect and high-$T_c$ superconductivity, but they
also turned out to influence the physical properties of metallic
1D and quasi-1D materials in a very profound way. In fact,
contrary to their higher dimensional relatives, which are almost
perfectly described by the Fermi liquid theory, the 1D metals
even constitute a special universality class of
Tomonaga-Luttinger liquids (TLL) \cite{haldane,luttinger}. Very
naturally, it is extremely important to have precise, reliable and
easily accessible methods to investigate electronic interactions.
The most useful of them involve quantities which are only present
in interacting systems such as, e.g., the high energy tails of
field emission spectra \cite{lea-gomer,gadzuk_plummer,komnik1}.

During the last two decades scanning tunneling microscopy and
atomic force microscopy as well as their descendants established
themselves as some of the most efficient techniques for the
investigation of low dimensional materials. It became possible to
image the electronic density of, e.g., carbon nanotubes, with
atomic resolution \cite{odomlieber,PhysRevB.61.2991}. In one of
the recent experiments \cite{leeeggert} even an explicit imaging
of Friedel oscillations (FOs) near the end of a single-wall carbon
nanotube (SWNT) has been successfully achieved, making it possible
to obtain the signatures of a TLL.

\begin{figure}[t]
  \begin{center}
  \includegraphics[width=0.9\columnwidth,draft=false]{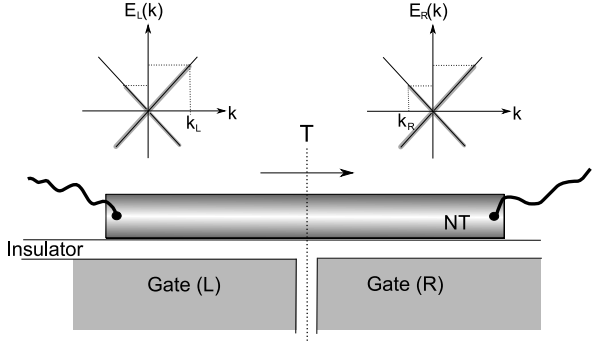}
  \end{center}
  \caption[]{Sketch of the system under investigation.}\label{fig1}
\end{figure}

The classical FO is nothing else but an electron density
modulation in vicinity of a hard wall potential \cite{mahan}. Due
to the full reflection at a boundary at $x=0$ a coherent
superposition of incoming and outgoing electron wavefunctions
results in a sin$(2k_F x)/x$ behavior of the correction to the
uniform electron density in the bulk (at $x \rightarrow\infty$).
The fundamental oscillation period then is defined by the Fermi
wave vector $k_F$ and is equal to $\pi/k_F$. Similar density
oscillations also emerge in the vicinity of any scatterer
introduced into a metal. Their form and the actual particle
density at the position of the scatterer of course depend on the
scattering potential. One could conclude that in the case of a
finite barrier transmission the modulation can change
qualitatively so that a beating pattern due to two harmonic
components would emerge as soon as the Fermi wave vectors on both
sides of the scatterer become different, $k_L \neq k_R$. A
straightforward calculation immediately shows that this is not the
case as long as electron-electron interactions are not taken into
account. Suppose we have a scatterer with reflection and
transmission amplitudes $r$ and $t$, respectively, placed at
$x=0$. Let $d_k$ and $a_k$ be the plane wave annihilation
operators for the electrons injected into the system at $x
\rightarrow \pm \infty$. Then the full field operator for
electrons on the left side of the scatterer is
\begin{equation}
\label{eq:LOperator}
 \psi_L(x) \!=\! \int \frac{d k}{2 \pi} \nu(k) \left[ \left( e^{i
 k x} - r e^{- i k x} \right) a_k + t^* e^{- i k x} d_k \right],
\end{equation}
where for simplicity we assume the band structure encoded in the
density of states $\nu(k)$ to be the same on both sides of the
scatterer. The particle density is then given by the average
$\langle \psi_L^\dag (x) \psi_L(x) \rangle$. Taking into account
that cross correlations of the type $\langle a^\dag_k \, d_{k'}
\rangle$ vanish, one realizes that the term proportional to $t^*
t$ is not $x$-dependent. That is why the density oscillations on
the (here) left side of the scatterer are not affected by the
electrons from the right side up to an overall uniform
$x$-independent shift. It is instructive to carry out the
calculation for a specific system. We use the model for the
conductance band of a SWNT: the electrons are assumed to be
one-dimensional with a dispersion relation which is the same on
both sides of the scatterer and is linear $E(k) = v_F k$, where
$v_F$ is the Fermi velocity (for simplicity we use units with
$\hbar = e = m_e = 1$ throughout), see Fig.~\ref{fig1}.
Different values of $k_F$ on both sides can be produced by
applying a finite bias voltage to the whole system complemented by
appropriate gating of the constituent subsystems. For convenience
we shall call the difference in the Fermi energies `voltage' $V$:
$\EFLR = v_F k_{L/R} = v_F k_F \pm V/2$. Then the expectation
values $\langle a_k^\dag a_{k'} \rangle = 2 \pi \delta_{k k'}
n_L(k)$ and $\langle d_k^\dag d_{k'} \rangle = 2 \pi \delta_{k k'}
n_R(k)$ contain the Fermi distribution functions $n_{L/R}$ with
Fermi energies $\EFLR$. The particle density $\rho_{L/R}$ is given
by the sum of three terms: (i) the full constant average density
$\rho^{(0)}=k_F/\pi$; (ii) a constant shift proportional to $V$ on
both sides given by
\begin{eqnarray}            \label{xindepdensity}
  \rho^{(V)}_{L/R} = \pm(1-T)V/(2v_F\pi) \, ,
\end{eqnarray}
the difference of which can be reworked into Landauer's residual
resistivity dipole strength $p=k_F V (1-T)/(2 \pi n)$, where $n$
is the density of the charge carriers
\cite{PhysRevB.43.6434,landauerold}; (iii) a space dependent
oscillating part
\begin{equation}
\label{eq:rhoOsz}
 \rho_{L/R}^{(osc)}(x) = -\frac{\sqrt{1-T}}{2\pi}\;
   \frac{\sin(2 x \, k_{L/R}-\phiR)+\sin(\phiR)}{x}
    \, .
\end{equation}
Here $T=t^* t$ is the transmission probability and $\phiR$ is the
phase shift for a reflected electron. The full constant average
density $\rho^{(0)}$ usually is compensated by the constant
homogeneous background charge of the ions of the underlying
lattice structure and must therefore be subtracted in the
interaction terms. Eqs.\ (\ref{xindepdensity}) and
(\ref{eq:rhoOsz}) show explicitly that $\rho_L$ $[\rho_R]$ only
depends on $k_L$ $[k_R]$ in the absence of interactions. Different
ways to measure the induced chemical potential oscillations are in
detail discussed in \cite{PhysRevB.40.3409}.

Now we want to address the effect of interactions on the FOs. We
consider a spin-independent pointlike density-density interaction
with strength $U$. In terms of the  Keldysh Green's functions
(GF) defined as $g^<_{\alpha L}(x,y,t) = - i \langle
\psi_\alpha(x,t) \psi_L(y,0) \rangle$ for the lesser GF and a
similar definition of the retarded GF, the first-order correction
to the particle density is given by
\begin{eqnarray}                    \label{correctiontorho}
 && \delta \rho_L(x) = 4 U \, \mbox{Im} \, \int d t
 \int_{-\infty}^0 \, d y \, \\ \nonumber
 &\times&
 \sum_{\alpha=R,L} \rho_\alpha(y) \,
  \, g^{ret}_{L \alpha} (x, \alpha y , -t) \, g^<_{\alpha L} (\alpha y, x, t)
 \, , \nonumber
\end{eqnarray}
with the prescription $\alpha=R,L=(\pm)$, which accounts for two
different subsystems: the left one for $y<0$ and its counterpart
on the right side, $y>0$. This correction has a simple physical
interpretation as it comes from the electron excursions from the
measuring point $x$ to some other coordinate $y$ where it gets
scattered on the local electron density $\rho_\alpha(y)$ (which
may be on either side of the barrier) and then propagates back to
the original $x$. The scattering process is crucial as it allows
for exchange of momenta. In this way the information about the
Fermi wave vector on the other side of the impurity is carried
back to the measuring point $x$. This effect is similar to
secondary tunnelling processes (Auger like scattering) responsible
for high-energy tails in field emission spectra of correlated
hosts \cite{lea-gomer,gadzuk_plummer,komnik1}. The GFs necessary
for the evaluation of (\ref{correctiontorho}) are

\begin{widetext}
 \begin{eqnarray}
  g_{LL}^{ret}(x,y,t) &=& - i
    \, \Theta(-t) \int_0^\infty\! \frac{d k}{\pi} \, e^{-i v_F k t } \,
   \left\{\cos[k(x-y)]  - \sqrt{1\!-\!T} \,
   \sin[k (x+y)-\varphi_r] \right\}\, ,
   \nonumber
\\
   g_{LL}^{<}(y,x,t)
   &=& -\int_0^\infty\!\frac{dk}{2\pi i} \left\{
   n_L \, e^{ik(y-x)} + [n_L (1\!-\!T) + n_R T] \, e^{-ik(y-x)} - 2 n_L
   \sqrt{1\!-\!T}\,\cos\left[
   k(y\!+\!x)\!-\!\varphi_r \right] \right\}e^{- i v_F k t}
   \nonumber \, ,
\\
  g_{LR}^{ret}(x,y,t)
  &=& - i \Theta(-t)
  \frac{\sqrt{T}}{\pi} \int_0^\infty d k\, e^{-i v_F k t} \,
  \sin\left[k(x+y) - \varphi_r \right]
  \nonumber \, ,
\\
   g_{RL}^{<}(y,x,t)
   &=&-\int_0^\infty\frac{dk}{2\pi} \left\{
   n_L \sqrt{T}\,e^{-ik(y+x)+i\varphi_r} - n_R \sqrt{T}\,e^{ik(y+x)-i \varphi_r}
   - \sqrt{T(1-T)} \left(n_L\!-\!n_R \right)\,e^{ik(x-y)}\right\}e^{-i v_F k t}
    \nonumber \, ,
\end{eqnarray}
\end{widetext}
where $\Theta$ denotes the unit step function. Note that these GFs
are exact in scattering strength since we are using the exact
scattering states. As we are looking for the secondary FO
(contrary to the primary one coming from the native $k_{L}$) it is
more convenient to perform the Fourier transform of the correction
to the density. Furthermore, we would like to concentrate on the
consequences of having two different Fermi levels in the system,
thus we only keep terms proportional to the difference of the
Fermi distribution functions $\Delta n=(n_L - n_R)$.

\begin{figure}[t]
   \begin{center}
   \includegraphics[width=0.6\columnwidth,draft=false]{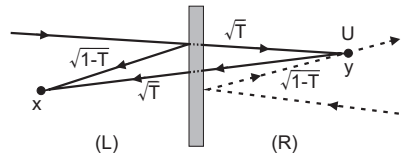}
   \end{center}
   \caption[]{Graphical representation of the processes contributing to the secondary
   Friedel oscillations, see text.}\label{fig3}
\end{figure}

\begin{figure}[t]
  \begin{center}
  \includegraphics[width=\columnwidth,draft=false]{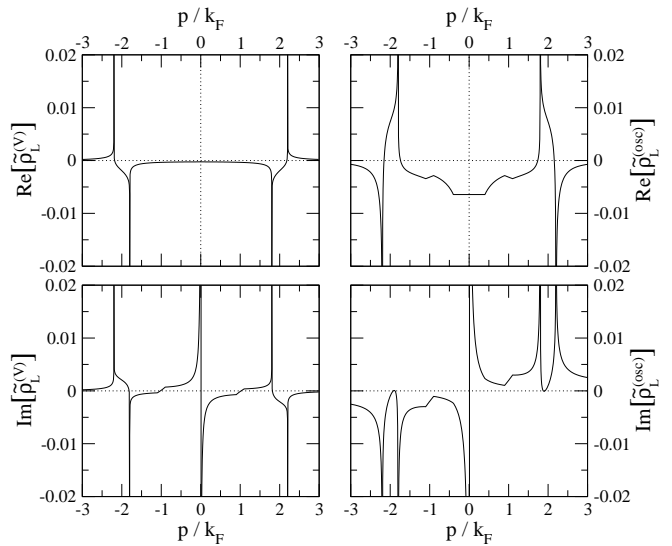}
  \end{center}
  \vspace*{-0.3cm}
  \caption[]{Real and imaginary parts of the density oscillations spectrum (arbitrary units)
  for $V=0.1 k_F^{-1}$.
  \emph{Left panels:}
  contributions from $\rho_{\alpha}^{(V)}$. \emph{Right panels:} contribution from the oscillating
  part of the density $\rho_\alpha^{(osc)}$.  }\label{fig2}
\end{figure}

The detailed evaluation of (\ref{correctiontorho}) reveals two
contributions to the first order density correction $\delta\rho$.
The first one comes from the mean shift $\rho^{(V)}_{L/R}$ of the
zeroth order density due to a finite $V$ and it is found to be
given by
\begin{eqnarray}
\label{eq:ErgILRneqRegReKVwT}
  \lefteqn{
  \delta \widetilde{\rho}_L^{(V)}(p)=\frac{T(1-T) V U}{4(\pi v_F)^2}
  }&&
  \\
&&
  \times\sum_{\xi=\pm}\frac{1+e^{-2i\xi\varphi_r}}{\xi p}\bigg\{\pi\,\Theta(\xi p+2 k_L)\Theta(-\xi p-2 k_R)
\nonumber\\
&&
  -i\xi\!\! \sum_{\alpha=R,L}\!\alpha
  \log\left|1+\frac{2k_\alpha}{\xi p}\right|
  \left[
  1-
  \Theta(\xi p+k_\alpha)(1+e^{2 i \xi \varphi_r}) \right]\!\!\bigg\}
\nonumber
\end{eqnarray}
The spectrum of the density profile shows up \emph{two} distinct
peaks at $p=\pm 2 k_{L/R}$ while in the noninteracting case there
would be only one native periodicity with wavelength $\pi / k_L$.
The peaks are in fact logarithmic singularities due to the
one-dimensional geometry of the system. By construction the signal
strength is proportional to the applied voltage. It vanishes for
both perfect transmission, when translational invariance of the
system is restored and no buildup of FOs is possible, and perfect
reflection, when the electrons on the one side of the barrier
cannot acquire information about the Fermi edge on the other side.
The prefactor $T(1-T)$ can indeed also be explained more
intuitively. Suppose we have a wave packet approaching the barrier
coming from the left reservoir. At the boundary a part of it is
reflected with an amplitude $\sim \sqrt{1-T}$. The transmitted
part which is $\sim \sqrt{T}$ propagates to the point $y$ where it
interacts with the wave packet coming from the right reservoir,
which on its way to $y$ is reflected from the boundary picking up
a prefactor $\sim \sqrt{1-T}$. The interaction process yields a
prefactor $U$ and the propagation back through the barrier gives
another $\sqrt{T}$, see Fig.~\ref{fig3}. Thus the full prefactor
is $(\sqrt{T} \, \sqrt{1-T})^2=T(1-T)$. For a representative set
of parameters the density spectra as given by
(\ref{eq:ErgILRneqRegReKVwT}) are plotted in the left panels of
Fig.~\ref{fig2}. Being transformed back into the coordinate space,
the two peaks result in a beating pattern of the density
oscillation as a function of distance from the impurity.

Besides these oscillations, originating from the finite (average)
coordinate-independent difference in chemical potentials, there
is, of course, a contribution from the oscillating part of the
density on the other side of the barrier. The physical motivation
and qualitative dependence on the transmission is the same as
discussed above, but corresponding expressions are rather long, so
we restrict ourselves to the plots shown in the right panels of
Fig.~\ref{fig2}. As before, distinct features arise around
\emph{both} doubled Fermi momenta $p=\pm 2 k_{L/R}$. We expect
that this interaction signature should be observable in dedicated
experiments, thereby giving important insights into the nature of
electronic correlations.

So far we treated the system exactly in the tunnelling
(scattering) amplitude and performed a weak interaction expansion.
In strictly 1D environments, at least in equilibrium, the electron
system belongs to the universality class of TLLs. In this case the
interactions have to be treated nonperturbatively. For the FOs at
an open boundary (at a hard wall) the density profile can be
calculated analytically and numerically for arbitrary interaction
strength
\cite{PhysRevLett.75.3505,PhysRevB.51.17827,PhysRevB.54.13597}. In
our case the situation is much more difficult as we must allow for
at least weak electron tunnelling at a barrier between two
half-infinite TLLs. It turns out, that such a situation indeed
describes the low-energy fixed point of a TLL with an impurity
\cite{PhysRevB.46.15233,PhysRevB.47.4631}. Therefore we adopt
this model and perform a calculation which is `dual' to the one
presented above, taking into account the interactions exactly
while performing a perturbation expansion in the tunnelling
amplitude.

\begin{figure}
  \begin{center}
  \includegraphics[width=\columnwidth,draft=false]{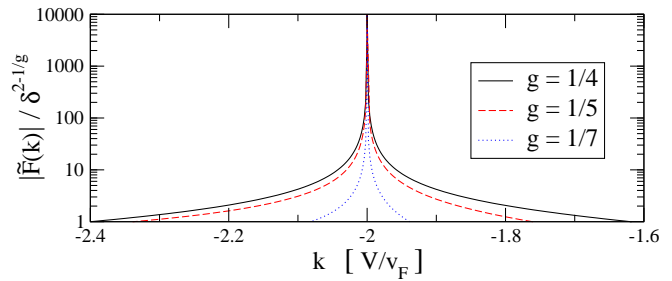}
  \end{center}
  \caption[]{The function $\widetilde{F}(k)$ as given by Eq.\
  (\ref{eq:FofKinLL}) plotted for different values of the interaction parameter $g$.
 }\label{fig4}
\end{figure}

The beating pattern in the density oscillations already shows up
in the inhomogeneous case (which is more convenient to handle),
when a noninteracting subsystem is coupled to a genuine TLL and
the FOs are measured in the interacting part 
The correction to the density profile is then found to be given by
\begin{eqnarray}            \label{TLLcorrection}
 \delta \rho(x) = \sum_{\zeta=\pm} e^{i \zeta (2 k_F + V/v_F) x}
 \, \delta \langle \psi_L^\dag(-\zeta x) \psi_L(\zeta x)
 \rangle \, .
\end{eqnarray}
The evaluation of the expectation value on the rhs is accomplished
with the help of the bosonization representation of the TLL
Hamiltonian with the interaction of the kind we already have used
in the first part of this Letter. The correlation strength
translates into the dimensionless interaction parameter
$g=1/\sqrt{1 + U/(\pi v_F)}$, with $0<g<1$ for repulsive
interaction. The correction to the density is
\begin{eqnarray}   \nonumber
 \delta \langle \psi_L^\dag(x) \psi_L(-x) \rangle = \frac{2 \gamma^2}{\pi
 v_F} \, \frac{a_0^{g+1/g}}{(\pi a_0)^2 \, |2 x|^g } \, e^{ i
 (\pi/2)
 \mbox{sign}(x)} \, F(x) \, ,
\end{eqnarray}
where $a_0$ is the lattice spacing of the underlying lattice
model. $F(x)$ is an oscillatory function, whose Fourier transform
can be written in the form
\begin{eqnarray}
\label{eq:FofKinLL}
 \widetilde{F}(k) &=&  \mbox{Re} \, \left\{
  -\int_0^\infty \, d \tau_+ \, \int_{-\tau_+}^{\tau_+} d \tau_- \, \left(
 \frac{- i }{\tau_- + i \delta} \right)^{1/g}
 \right. \nonumber \\
 &\times& \left. \frac{\tau_+ + \tau_-/2}{2 \tau_+ + i \delta} \, e^{ i k \tau_+ + i
 (k/2 + V/v_F) \tau_-} \,   \right. \nonumber \\
&-& \left. \sum_{\xi = \pm} \, \xi \int_0^\infty \, d \tau_+ \,
\int_{0}^{\tau_+} d \tau_- \, \left(
 \frac{\xi i }{\tau_- - i \xi \delta} \right)^{1/g}
\right. \nonumber \\
 &\times& \left. \frac{\tau_+ + \xi \tau_-/2}{2 \tau_+ + i \delta} \, e^{ i \xi k \tau_+ + i
 ( k/2 + V/v_F)  \tau_- } \,
 \right\}  \, .
\end{eqnarray}
As $1/g>1$ for repulsive interactions it is the
$\tau_-$--dependent parts in the brackets which dominate the
behaviour of $\widetilde{F}(k)$. A stationary phase argument with
respect to the $\tau_-$ integration yields the condition $k/2 +
V/v_F=0$ for the maximum of $\widetilde{F}(k)$. This is confirmed
by the numerical evaluation of (\ref{eq:FofKinLL}) shown in Fig.\
\ref{fig4}, where $\widetilde{F}(k)$ is plotted for different
values of the interaction parameter $g$. Therefore we conclude
that the correction (\ref{TLLcorrection}) shows up oscillations
with period $\pi/k_R$ in addition to the native one with $\pi/k_L$
just like in the perturbative interaction calculation.
Interestingly, the stronger the interactions the more pronounced
are these newly arising oscillations. Similar effects in a
slightly different setup have been reported in
Ref.~\cite{PhysRevLett.77.538}.

To conclude, we have discussed the density profile of Friedel
oscillations in a 1D metal with different Fermi wave vectors on
either side of an imbedded impurity. Contrary to the naive
expectation in the absence of interactions, the density
oscillations possess only one periodicity originating from the
local value of $k_F$ even when the impurity has a finite
transmission coefficient. The situation changes dramatically in
interacting systems, where the spectral function acquires an
additional peak, which is a result of the interplay of
interactions and the Landauer dipole. We have shown this in two
different situations by first taking into account interactions
perturbatively and using the exact scattering states. In a second
step we have considered the dual situation of a weak link between
two TLLs where the electronic correlations have been taken into
account with the help of bosonization technique. We are convinced
that the very fact, that the beating pattern owes its existence to
the interactions has the potential to make measurements of Friedel
oscillations (which by now have become unproblematic) an important
tool to study electronic correlations in 1D metals and carbon
nanotubes in future experiments.

The authors would like to thank H.~Grabert and R.~Egger for many
interesting discussions. A.K. is supported by the DFG grant No. KO
2235/2 (Germany).
\bibliography{NEFriedel}

\end{document}